\documentclass[conference]{IEEEtran}
\IEEEoverridecommandlockouts
\usepackage{amsmath,amssymb,amsfonts}
\usepackage{amsthm}
\usepackage{float}
\usepackage{graphicx}
\usepackage{textcomp}
\usepackage{threeparttable}
\usepackage{optidef}
\usepackage{xcolor}
\usepackage{pifont}
\usepackage{makecell}
\usepackage{algorithm}
\usepackage{algpseudocode}
\usepackage{hyperref}
\hypersetup{hidelinks}

\def\BibTeX{{\rm B\kern-.05em{\sc i\kern-.025em b}\kern-.08em
    T\kern-.1667em\lower.7ex\hbox{E}\kern-.125emX}}
    
\usepackage[
backend=biber,
style=ieee,
sorting=none
]{biblatex}

\begin{document}


\title{On the airspace complexity metrics for predecessor-follower operations\\
}

\author{\IEEEauthorblockN{Lucas Souza e Silva and Luis Rodrigues}
\IEEEauthorblockA{\textit{Department of Electrical and Computer Engineering} \\
\textit{Concordia University}\\
Montreal, Canada 
}
}
\maketitle

\begin{abstract}

This technical note proposes a novel airspace complexity metric that quantifies the air traffic controller workload and coordination effort for pairwise predecessor-follower aircraft operations in cruise. The pairwise dynamic workload (PDW) is proposed as a continuous function that depends on the relevant parameters of these operations, such as the aircraft separation and separation rate. A comparison of this metric with the dynamic density (DD) shows that it is capable of continuously evaluating the variation of airspace complexity over time and monitoring the aircraft parameters that might lead to conflicts. This metric can be used to support the implementation of autonomous and supervised aircraft procedures, to achieve a more structured and coordinated airspace.
\end{abstract}


\section{Introduction}\label{Intro}

The airspace is becoming more complex due to the growing demand for both domestic and international travel. As a result, the augmented airspace complexity imposes greater strain on the efficiency of air traffic management (ATM). This strain is particularly intense on the air traffic control operators (ATCOs), who must perform high-pressure, time-sensitive tasks that call for sustained attention over extended work shifts \cite{NYT}. Moreover, new transportation technologies, such as advanced/urban air mobility (AAM/UAM) \cite{FAA_Conops}, require integration with the traditional air traffic control (ATC) environment. Thus, achieving a harmonized and coordinated air traffic management becomes increasingly challenging. One approach to alleviate these challenges is the use of structured and collaborative aircraft procedures. Several of these procedures can be characterized as platoons of pairwise predecessor–follower operations, in which safe separation between aircraft is maintained throughout the flight. Examples include aircraft sequencing based on scheduling in terminal areas and air mobility vehicles operating within designated air corridors. As the complexity of the airspace continues to grow, these operations necessitate automated or supervised ATC strategies to ensure their effective management. Several metrics have been proposed in the literature to quantify airspace complexity, depending on the type of operations of interest. For pairwise aircraft operating in cruise, it is essential to consider their horizontal separation and the rate at which this separation changes over time. References \cite{Pawlak1996, Chatterji2001, Laudeman1998} propose different complexity metrics and analyses that account for horizontal separation and separation rate. However, \cite{Pawlak1996} and \cite{Chatterji2001} do not provide a closed-form relation among these parameters, while \cite{Laudeman1998} only considers the number of aircraft in a sector that changed speed or are within unsafe distances from others, without capturing their continuous evolution over time. To address these limitations, this technical note introduces a novel pairwise dynamic workload (PDW) metric for predecessor–follower aircraft operations. The main contribution is the formulation of the PDW as a dynamic airspace complexity indicator that systematically captures the relevant operational variables governing pairwise interactions. By explicitly linking aircraft dynamics with workload-related airspace factors, the proposed metric provides a structured and analytically tractable framework for assessing complexity in pairwise predecessor–follower configurations.

The remaining of this technical note is structured as follows. Section \ref{sec_Complexity_Metric} defines the PDW metric in the context of pairwise aircraft operations in cruise. Section \ref{sec_Results} shows the comparison of the PDW with the dynamic density (DD) metric in simulated scenarios and section \ref{sec_Conc} concludes this note.

\section{Pairwise dynamic workload (PDW)} \label{sec_Complexity_Metric}
To dynamically describe the variation of the airspace complexity associated to pairwise predecessor-follower aircraft operations in cruise, it is assumed that both aircraft are flying at the same altitude. We propose a pairwise dynamic workload (PDW) function $\chi_{i,i-1}$ that satisfies the following conditions:

\begin{itemize}
    \item The function $\chi_{i,i-1}$ is continuous and strictly positive for all $t \in [0,t_f]$, where $t_f$ denotes the duration of the pairwise operation.
    
    \item The value of $\chi_{i,i-1}$ increases as the inter-aircraft separation $d_{i,i-1}$ decreases, and decreases when $d_{i,i-1} $increases.
    
    \item We define the aircraft separation rate as $\dot{d}_{i,i-1} = v_{i-1}-v_i$, where $v_{i-1}$ and $v_i$ are the predecessor and the follower aircraft airspeeds, respectively. An increase in the separation rate $\dot{d}_{i,i-1}$ should increase the derivative of $\chi_{i,i-1}$ with respect to $d_{i,i-1}$. Conversely, a decrease in $\dot{d}_{i,i-1}$ should reduce this derivative.
\end{itemize}

\textit{Definition 1 (Pairwise dynamic workload):} Given the separation $d_{i,i-1}$ and separation rate $\dot{d}_{i,i-1}$ of two aircraft in predecessor-follower operations of duration $t_f$ and given the maximum separation rate $\dot{d}_{max} > 0$, with $\dot{d}_{max} > |\dot{d}_{i,i-1}|$ $\forall$ $t \in [0,t_f]$, the proposed pairwise dynamic workload (PDW) function $\chi_{i,i-1}$ is
\begin{equation}
    \chi_{i,i-1} = \frac{1}{d_{i,i-1}} \biggl(1-\frac{\dot{d}_{i,i-1}}{\dot{d}_{max}} \biggl) \label{eq_PDW}
\end{equation}

Note that $\chi_{i,i-1} > 0$ for any finite  $d_{i,i-1}$, and it is inversely proportional to the inter-aircraft separation, as smaller separations require increased attention and workload from the ATCOs. This behavior is confirmed by the derivative in \eqref{eq_partialX_d}, where the numerator is always positive, which makes the derivative of $\chi_{i,i-1}$ with respect to $d_{i,i-1}$ always negative.
\begin{equation}
    \frac{\partial \chi_{i,i-1}}{\partial d_{i,i-1}} = -\frac{1-\frac{\dot{d}_{i,i-1}}{\dot{d}_{max}}}{d_{i,i-1}^2} \label{eq_partialX_d}
\end{equation}

Moreover, a negative closure rate $\dot{d}_{i,i-1}$ indicates that the aircraft are converging, thereby increasing the coordination workload for the ATCOs and contributing to greater airspace complexity. In contrast, a positive closure rate implies that the aircraft are diverging, which reduces the controller's workload and consequently lowers the associated complexity. This interpretation can be confirmed by inspecting the computed derivative in \eqref{eq_partialX_ddot}. The derivative of $\chi_{i,i-1}$ with respect to $\dot{d}_{i,i-1}$ is always negative, given that the denominator in \eqref{eq_partialX_ddot} is always positive.
\begin{equation}
     \frac{\partial \chi_{i,i-1}}{\partial \dot{d}_{i,i-1}} = -\frac{1}{d_{i,i-1}\dot{d}_{max}} \label{eq_partialX_ddot}
\end{equation}

\section{Simulation results}\label{sec_Results}

Simulations were conducted in MATLAB on a laptop with a configuration of 16 GB RAM and an 11$^{th}$ Gen Intel$^R$ Core$^{TM}$ i5-1135G7 2.40GHz CPU. A comparison of the proposed PDW with other airspace complexity metrics from the literature is shown in Table \ref{table_matric_comparison}.

\begin{table} [hbt]
   \caption{Comparison of Airspace Complexity Metrics for Pairwise Aircraft Operations}\label{table_matric_comparison}
   \centering
   \begin{tabular}{|c| c c c c|}
   \hline
   \textbf{Parameter} & \textbf{\cite{Pawlak1996}} & \textbf{\cite{Chatterji2001}} & \textbf{\cite{Laudeman1998}} & \textbf{PDW} \\
   \hline
   Horizontal Separation & \checkmark & \checkmark & \checkmark & \checkmark \\
   Separation Rate & \checkmark & \checkmark & \checkmark & \checkmark \\
   Closed-form & \ding{55} & \ding{55} & \checkmark & \checkmark \\
   Temporal evolution of parameters & \ding{55} & \checkmark & \ding{55} & \checkmark \\
     \hline
     \end{tabular}
\end{table}

The metrics proposed in \cite{Pawlak1996} and \cite{Chatterji2001} do not have a closed-form equation that relates the aircraft separation and separation rate. For this reason, the proposed PDW metric will be evaluated alongside the dynamic density (DD) measure introduced in \cite{Laudeman1998}. The comparison is conducted for two distinct scenarios involving a pairwise predecessor–follower procedure over a 60-minute duration. In these scenarios, $v_0(t)$ and $v_1(t)$ denote the airspeed of the predecessor and the follower aircraft, respectively.  The profiles and values of $v_0(t)$ and $v_1(t)$ were selected to generate multiple airspeed variations while maintaining a constant sign of $\dot{d}_{i,i-1}$. The two scenarios are described below.

\begin{enumerate}
    \item Scenario 1, aircraft converging: the predecessor and follower aircraft airspeeds are, respectively,

    \begin{equation}
       v_0(t) =  \begin{cases}
           600 \ \mathrm{km/h}, \text{for} \ t \in [0,10), [20,30), [40,50) \\
           650 \ \mathrm{km/h}, \text{for} \ t \in [10,20), [30,40), [50,60)           
       \end{cases}
    \end{equation}

    \begin{equation}
        v_1(t) = \begin{cases}
            700 \ \mathrm{km/h}, \text{for} \ t \in [0,30) \\
            670 \ \mathrm{km/h}, \text{for} \ t \in [30,60)
        \end{cases}
    \end{equation}

    \item Scenario 2, aircraft diverging:  the predecessor and follower aircraft airspeeds are, respectively,
    
        \begin{equation}
        v_0(t) = \begin{cases}
            700 \ \mathrm{km/h}, \text{for} \ t \in [0,30) \\
            670 \ \mathrm{km/h}, \text{for} \ t \in [30,60)
        \end{cases}
    \end{equation}

        \begin{equation}
       v_1(t) =  \begin{cases}
           600 \ \mathrm{km/h}, \text{for} \ t \in [0,10), [20,30), [40,50) \\
           650 \ \mathrm{km/h}, \text{for} \ t \in [10,20), [30,40), [50,60)           
       \end{cases}
    \end{equation}

\end{enumerate}

\begin{figure}[h!]
\centerline{\includegraphics[scale=0.52]{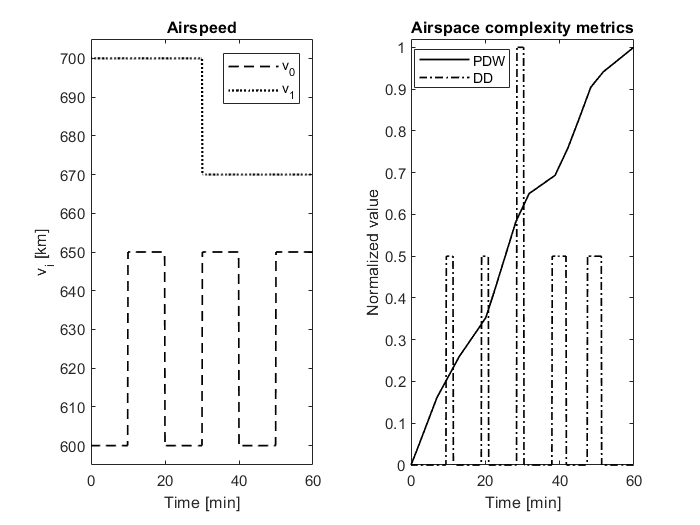}}
\caption{Comparison of PDW and DD, scenario 1}
\label{fig_complexity_met_comp_1}
\end{figure}

\begin{figure}[h!]
\centerline{\includegraphics[scale=0.52]{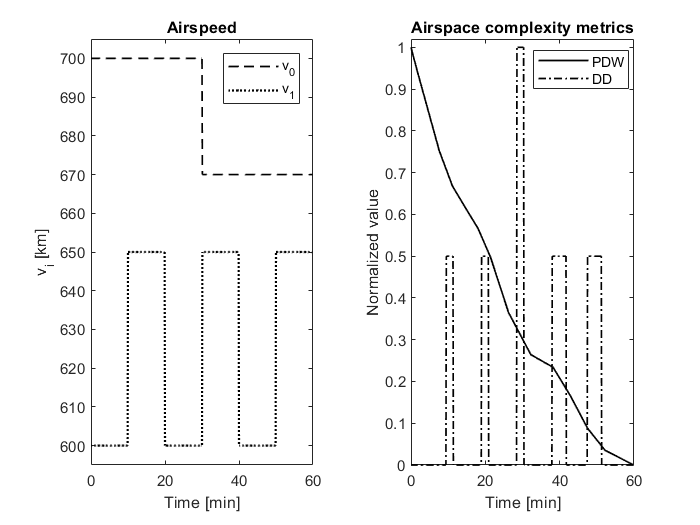}}
\caption{Comparison of PDW and DD, scenario 2}
\label{fig_complexity_met_comp_2}
\end{figure}

Both PDW and DD metrics were computed for the two scenarios and scaled using min–max normalization, yielding values between 0 (minimum) and 1 (maximum), as shown in Figure \ref{fig_complexity_met_comp_1} and Figure \ref{fig_complexity_met_comp_2}. The left plots present the airspeed profiles for each scenario, while the right plots depict the evolution of airspace complexity metrics over time. In Scenario 1, $v_1(t) > v_0(t)$, leading to the convergence of the two aircraft, whereas in Scenario 2, $v_1(t) < v_0(t)$, resulting in increasing separation. In both scenarios, DD produced identical values, since it only accounts for the number of aircraft that changed airspeed within each 2-min interval. In contrast, the PDW metric captured the continuous influence of both the separation and its rate of change on airspace complexity, increasing when the aircraft are converging and decreasing when they are diverging. Furthermore, the rate of separation between the aircraft was observed to affect the slope of the PDW curve, as higher separation rates lead to a steeper gradient. Unlike DD, PDW reflects gradual changes and transient behaviors in pairwise aircraft interactions. Moreover, PDW highlights situations where conflicts may be emerging even before DD detects them.

\section{Conclusions}\label{sec_Conc}
This technical note introduces the pairwise dynamic workload (PDW) as an airspace complexity metric. To capture the continuous impact of pairwise operations on airspace structure and controller workload, the PDW is a defined as a continuous function that depends on the aircraft separation and separation rate. The simulation results show that the PDW models incremental variations in the airspace parameters that configure pairwise operations and may anticipate potential conflicts faster when compared to the dynamic density metric. This metric may support the air traffic management by providing a dynamic and continuous complexity metric to aid the integration of autonomous or supervised aircraft procedures to achieve a more structured and coordinated airspace.

\printbibliography

\end{document}